\documentclass[a4paper, conference]{IEEEtran}
\IEEEoverridecommandlockouts
\usepackage{cite}
\usepackage{amsmath,amssymb,amsfonts}
\usepackage{algorithmic}
\usepackage{graphicx}
\usepackage{textcomp}
\usepackage{xcolor}
\usepackage{tabularray}

\usepackage{background}
\usepackage{xcolor}
\usepackage{hyperref}

\hypersetup{
  pdfborder={0 0 0}, 
}

\def\BibTeX{{\rm B\kern-.05em{\sc i\kern-.025em b}\kern-.08em
    T\kern-.1667em\lower.7ex\hbox{E}\kern-.125emX}}

\backgroundsetup{
  scale=1,
  color=black,
  opacity=1,
  angle=0,
  position=current page.south,
  vshift=10pt,
  contents={\textcolor{red}{© 2023. For personal use only. Published version under doi:  \href{https://doi.org/10.23919/SoftCOM58365.2023.10271664}{10.23919/SoftCOM58365.2023.10271664}. }}
}

\begin{document}

\title{An Overview of Machine Learning-Enabled Network Softwarization for the Internet of Things}

\author{\IEEEauthorblockN{Mohamed Ali Zormati}
\IEEEauthorblockA{\textit{Heudiasyc UMR 7253} \\
\textit{Sorbonne University Alliance}\\
\textit{Université de Technologie de Compiègne (UTC)}\\
Compiègne, France \\
zormati@ieee.org}
\and
\IEEEauthorblockN{Hicham Lakhlef}
\IEEEauthorblockA{\textit{Heudiasyc UMR 7253} \\
\textit{Sorbonne University Alliance}\\
\textit{Université de Technologie de Compiègne (UTC)}\\
Compiègne, France \\
hicham.lakhlef@hds.utc.fr}
}

\maketitle

\begin{abstract}
The Internet of Things (IoT) has evolved from a novel technology to an integral part of our everyday lives. It encompasses a multitude of heterogeneous devices that collect valuable data through various sensors. The sheer volume of these interconnected devices poses significant challenges as IoT provides complex network services with diverse requirements on a shared infrastructure. Network softwarization could help address these issues as it has emerged as a paradigm that enhances traditional networking by decoupling hardware from software and leveraging enabling technologies such as Software Defined Networking (SDN) and Network Function Virtualization (NFV). In networking, Machine Learning (ML) has demonstrated impressive results across multiple domains. By smoothly integrating with network softwarization, ML plays a pivotal role in building efficient and intelligent IoT networks. This paper explores the fundamentals of IoT, network softwarization, and ML, while reviewing the latest advances in ML-enabled network softwarization for IoT.
\end{abstract}


\section{Introduction}
The Internet of Things (IoT) is a promising technology that creates connectivity of anything from anywhere at any time \cite{b3}. It impacts all areas of human life as it is applied in diverse fields \cite{b16}. It is growing exponentially \cite{b2} as the number of devices is expected to exceed 30 billion by 2025. The huge number of interconnected heterogeneous devices poses new challenges to IoT networks as these devices generate a massive and increasing amount of data \cite{b16}  \cite{b6}. Managing such a heterogeneous network poses technical challenges, and to overcome these obstacles, IoT networks should be combined with other technologies, such as network softwarization \cite{b5}.

Network softwarization is a novel key enabling technology to meet IoT requirements for flexibility \cite{b3}. It combines both Software Defined Networking (SDN) and Network Function Virtualization (NFV) and aims to transform the communication process and system components from legacy network devices to general purpose devices \cite{b7}, enabling a wide range of services, while ensuring remarkably low capital and operational expenditures (i.e., CAPEX and OPEX) \cite{b5}. 

SDN is a revolutionary networking technology that decouples the data plane (i.e., forwarding devices) from the control plane (i.e., network control logic) to provide a centralized global view of the network \cite{b1}. In this way, SDN enables programmable networks that are easier to manage, configure and control \cite{b3} \cite{b5}. As a key enabler for network softwarization, SDN has rapidly grown together with NFV \cite{b7}.

NFV decouples various network functions from physical network devices (i.e., proprietary hardware) and executes them using virtualization technologies \cite{b3} \cite{b11}, and thus, simplifying resource management, and provisioning of network functions \cite{b5}. Network functions are abstracted as logical entities and could be deployed on-demand as Virtualized Network Functions (VNFs) \cite{b11}.

Although they do not require each other to be implemented, SDN and NFV are complementary and mutually reinforcing \cite{b29}. However, for an effective implementation, many challenges need to be fully addressed (e.g., security concerns, resource utilization) \cite{b3} \cite{b13}. Machine Learning (ML) can potentially be used to address these issues.

ML is a category of Artificial Intelligence (AI) based on intelligence that can learn from data, make decisions, identify patterns, and perform various actions with less human intervention \cite{b7}. ML techniques thus allow networks to learn from experience to make them more robust against vulnerabilities and failures and to improve performance.

Although many research efforts have been made to investigate these three promising technologies and their combination, to the best of our knowledge, no existing work has actually focused on ML-enabled network softwarization for IoT networks. Therefore, our work is intended to fill this gap.

The rest of the paper is organized as follows. In Sec. II, we introduce the background of IoT, network softwarization, and ML techniques. In Sec. III, we review how ML algorithms are applied in network softwarization and how SDN and NFV techniques are applied in IoT networks, with a focus on intelligence. Challenges and future directions are discussed in Sec. IV. We conclude this review in Sec. V.

\section{Background}
In this section, we present the main concepts of IoT, network softwarization with an emphasis on SDN and NFV technologies, and ML techniques.

\subsection{Internet of Things (IoT)}

The IoT connects billions of devices (i.e., things) to provide a variety of services by ensuring connectivity between devices and the Internet. IoT not only connects objects, but also focuses on optimizing and automating traditional systems, as demonstrated by the various IoT applications (e.g., smart cities, smart vehicles, healthcare)\cite{b15}.

Several architectures have been proposed for IoT \cite{b30}. According to \cite{b15}, most of the proposals consist of three layers: perception layer (contains the objects used for sensing), network layer (contains the network devices and communication technologies), and application layer (contains the IoT services and applications).


The complexity of IoT is mainly due to the fact that a large number of things are connected to the Internet and communicate with each other, while being governed by different protocols and models \cite{b15}. Moreover, IoT faces several challenges including scalability, standardization, and security \cite{b30} \cite{b25}. To address these issues and challenges, and to meet the requirements of IoT, many research works propose network softwarization as a key enabling technology for IoT networks.

\subsection{Network softwarization}
Network softwarization is an emerging approach to transform the traditional networks to the new trending technologies such as programmable network and virtualization. By leveraging the software features such as flexibility, it aims to deliver new services and functions and deploy them in an efficient way \cite{b15}. The goal is to softwarize the network and transform it into an open ecosystem that dissociates the hardware and software involved in the network. This is expected to provide benefits in terms of advanced services, advanced networking capabilities, combined with network development and maintenance capabilities \cite{b24}.

Both SDN and NFV technologies play an important role in enabling network softwarization and increasing programmability \cite{b15}, along with other emerging technologies (e.g., cloud computing, edge computing, network slicing) \cite{b24}.

\subsubsection{Software Defined Networking (SDN)}~\par
Over the past decade, a new wave of innovation in networking has emerged thanks to the SDN paradigm \cite{b7}. SDN enables the decoupling of the network control plane and data plane. It provides a set of Application Programming Interfaces (APIs) that can implement common network services for business purposes \cite{b12}.

By decoupling the control plane and data plane, SDN improves the network architecture and eliminates its hierarchy \cite{b15}. Network devices (e.g. routers, switches, access points) become forwarding devices. The centralized controller is responsible for managing and controlling all network functions and can dynamically program the network \cite{b14}. The centralized architecture of SDN provides a faster overview of the network status and much smoother programmability and updates \cite{b7} and has the potential to increase network flexibility and performance \cite{b1}.

SDN faces several challenges that need to be addressed for effective implementation. The most important ones are scalability, security, reliability, and interoperability \cite{b7}.

The architecture of SDN consists of three planes that can communicate through APIs \cite{b7} \cite{b15} \cite{b14}: data plane (consists of the set of physical or virtual network devices), control plane (is responsible for making networking decisions, its main component is the logically centralized SDN controller), and application plane (consists of business applications that satisfy user requirements). 

To facilitate communication between different layers and multiple components, SDN defines four types of interfaces \cite{b15}. Southbound interface is used to ensure communication between data plane and control plane. Northbound interface is used as an interface between the control plane and the application plane. To ensure coordination between multiple SDN controllers, westbound and eastbound interfaces can be used \cite{b16}.

\subsubsection{Network Function Virtualization (NFV)}~\par
SDN has rapidly grown together with the NFV concept, as they combined forces to boost emergent networking applications \cite{b5} \cite{b7}. NFV is a key enabling technology to deliver on-demand network services.

NFV transfers the network functions from specific hardware to software virtualized platform. In that, the network devices are hosted in general-purpose hardware using virtualization technologies \cite{b15}. Thus, network functions become Virtual Network Functions (VNFs) that are implemented on multiple components over multiple virtual machines.

The NFV architecture is composed of three main layers: Network Function Virtualization Infrastructure (NFVI) layer (provides the infrastructure of general-purpose physical devices and virtualization environment),  NFV Management and Orchestration (MANO) layer (is  responsible for managing all virtualization processes of NFV), and Virtual Network Function (VNF) layer (is responsible for providing functionalities and services) \cite{b15}.

Admittedly, NFV is still in its early stages, as many key challenges need to be thoroughly addressed \cite{b28}, and one of the most important challenges is how to optimally allocate virtual resources to network services \cite{b3}.

\subsection{Machine Learning (ML)}
Thanks to network softwarization, the application of Artificial Intelligence (AI) and ML to networking is easier to implement nowadays \cite{b7}. ML is a branch of AI that defines any computational method in which the results of previous events or decisions are used to improve predictions or decisions \cite{b4}, allowing systems to improve themselves without being explicitly programmed.

ML is a very large field whose methods have been classified into several categories. We propose a classification approach based on the type of learning involved.

\subsubsection{Supervised Learning (SL)}
It is based on discovering the unknown function that relates the input and output spaces from input-output pairs. The most commonly used supervised methods are (although some of them can be trained unsupervised or using a reinforcement learning strategy): k-Nearest Neighbor (k-NN), Linear Regression (LR), Markov Decision Process (MDP), Random Forests (RF), and Artificial Neural Network (ANN). We also have to introduce the well known Deep Neural Network (DNN). Commonly referred to as Deep Learning (DL), it is a category of ANNs that includes network architectures that have in common the high number of interconnected layers. Convolutional Neural Network (CNN) and Recurrent Neural Network (RNN) are two main types of DNNs.

\subsubsection{Unsupervised Learning (UL)}
It searches for patterns in unlabeled datasets. Unlike SL, human supervision disappears due to the lack of pre-labeled input-output pairs. Unsupervised methods self-infer relationships between variables according to features (e.g. correlations). The UL techniques are widely used in clustering and data aggregation and proceed by clustering sample data into different groups according to the similarity between them. The most common UL methods are \cite{b7} \cite{b14}: k-Means, Gaussian Mixture Model (GMM), and Self-Organizing Map (SOM).

\subsubsection{Reinforcement Learning (RL)}
It is an ML paradigm designed to teach an agent (i.e., the learning entity) to make local decisions and take actions to maximize a cumulative long-term reward (via feedback from the environment) \cite{b14}. Deep Reinforcement Learning (DRL) is a subset of RL that combines DNNs with RL models, exploiting the powerful function approximation property of DNNs. In RL, the agent monitors the state of the environment, and at each time step chooses an action, receives an immediate reward indicating how good or bad the action is, and transitions to the next state. The agent's goal is to learn the optimal behavioral policy to maximize the expected long-term reward \cite{b18}. Q-learning is a model-free reinforcement learning method to teach an agent an action policy according to the state and the observations from the environment. Today, it is the basis for existing RL models. As RL evolved into DRL, Q-learning evolved into Deep Q-learning \cite{b7} by replacing the Markov Decision Support (MDS) framework with DNN, thus solving the problem of multiple states and massive data.

\subsubsection{Federated Learning (FL)}
Is is one of the most attractive techniques that allows a heterogeneous set of devices to train an ML model without sharing their raw data \cite{b36} \cite{b34} \cite{b35}. While improving privacy and communication efficiency, FL also leverages massive, distributed data and computational resources in constrained networks (e.g., IoT networks) \cite{b37}. For example, in FL, only learning model updates are transmitted between end devices and the FL aggregation server \cite{b38}. The FL process generally consists of three steps \cite{b39} \cite{b40}: training task and global model initiation, local model update, and global aggregation.

The selection of the best ML method mainly depends on the dataset type, the dataset size, and the problem type.

\section{ML-Enabled Network Softwarization for IoT}
In IoT networks, the large number of heterogeneous devices poses enormous challenges. This opens up many research areas to build dynamic IoT networks and overcome the limitations. Network softwarization and ML techniques are enabling technologies that should be considered to address the above issues. In this section, we first present the existing applications of network softwarization techniques in IoT. Then, we highlight the existing work on intelligent network softwarization. Finally, we review the state of the art in combining ML, SDN, and NFV with IoT networks.

\subsection{Network softwarization for IoT}
In IoT, network scalability and flexibility are critical. SDN and NFV are suitable network softwarization technologies to enable these functionalities. This topic has received a lot of attention from the research community. In the following, we present some relevant papers on this topic.

In \cite{b17}, the authors address the challenge of bringing SDN architecture (which generates an overhead that can affect overall network performance) to IoT networks (specifically, IEEE 802.15.4 low-power wireless networks). They present a lightweight SDN framework, \(\mu\)\textit{SDN}, and evaluate it in terms of latency, energy, and packet delivery. Simulation results show that the proposed solution maintains scalability compared to a conventional IEEE 802.15.4 network with Routing Protocol for Low-Power and Lossy Networks (RPL) routing. Note that the authors do not apply NFV techniques in conjunction with SDN, even though they note that SDN enables NFV.

In \cite{b11}, the authors propose an energy-aware SDN and NFV-enabled architecture for IoT. The authors argue that data aggregation (which can be defined as an NFV instance and dynamically deployed to the required IoT nodes) can further reduce network traffic and improve overall resource utilization. After detailing the proposal, the authors proceed to a simulation using \textit{Cooja}. Extensive evaluation confirms that the proposed solution outperforms its counterparts in terms of energy consumption and Packet Delivery Ratio (PDR). It should be noted that the proposed solution is based on the \(\mu\)\textit{SDN} as proposed by Baddeley et al. in \cite{b17}.

In \cite{b6}, the authors propose a new mechanism for load balancing routing and virtualization through SDN for IoT. The authors recall that load balancing has four main goals: resource utilization, Quality of Service (QoS), resilience, and scalability. By directly monitoring the link load information and network operation status, the \textit{OpenFlow} protocol is used to determine the load balancing routing for each flow in various IoT applications. After introducing the proposed algorithm, the authors evaluate its efficiency through simulation using \textit{NS2}, with the simulation results supporting the findings. Note that for this solution, it is assumed that there are no mobile nodes in the network, and thus the topology remains constant until a node loses its energy and leaves the network. However, IoT networks (e.g., sensor networks) are often dynamic with mobile nodes.

In conclusion, nowadays a great effort is being made to fully softwarize the IoT network through the SDN and NFV technologies, which is confirmed by the reviewed papers, thus addressing the main challenges of the IoT (e.g., energy efficiency, security, etc.). However, we point out that these references do not include intelligence in the solution. Intelligence, via ML techniques, is certainly a key feature to ensure an efficient network softwarization for IoT.

\subsection{ML-based network softwarization}
Nowadays, ML techniques are a driving force of several domains, and it is certainly a booster for network softwarization. Therefore, it is quite normal that academia has focused on the topic of enabling ML for SDN and NFV in recent years, as presented below.

In \cite{b18}, the authors propose an RL-based algorithm for selecting an appropriate path for a Service Function Chain (SFC - i.e., a sequential chain of VNFs) in an SDN- and NFV-enabled network. The proposed solution selects an appropriate path depending on the network conditions, thus ensuring an efficient service chaining environment. The selection method is based on Q-learning. The authors chose RL techniques because they believe that it is the most appropriate ML technique for decision making problems.  The reward depends on CPU and bandwidth usage.  The authors implement a simulator with \textit{Java} and evaluate the performance of their proposal in comparison with the Greedy method. The solution outperforms its counterpart by taking network conditions into account. However, we note that the authors did not consider constrained networks (e.g., IoT networks) and thus did not consider an energy-related metric to evaluate energy efficiency.

In \cite{b21}, the authors study a DRL-based framework for online end-user service provisioning in an NFV-enabled network. They formulate an optimization problem aimed at minimizing the cost of network resource utilization and propose a Deep Q-network to solve the aforementioned optimization problem. It is interesting to note that the authors evaluate the computational complexity of the proposed method. To evaluate the performance of the proposed method, the authors consider different baselines. The results obtained for different ranges of parameters show the effectiveness of the framework.

In summary, many research efforts have been made to merge ML with network softwarization and mainly SDN and NFV, but we note that authors usually consider SDN as the main key enabler for network softwarization and omit NFV. We also note that even if some works consider different networks (e.g., data centers), they omit IoT networks, which are continuously growing and require efficient intelligent network softwarization techniques.

\subsection{Intelligent IoT network softwarization}
ML plays an essential role in creating smarter IoT networks, as it has shown remarkable results in various domains. As the architecture of network softwarization enables the integration of ML, we strongly believe that we will soon see a fusion of ML and network softwarization techniques with IoT. In the following, we present the most recent and relevant works, even though there is a small number of publications on the topic.

In \cite{b8}, the authors propose a DRL energy efficient task assignment and scheduling in SDN-based IoT network. It should be recalled that task scheduling is challenging as it often manifests as a difficult online decision making. They strive to minimize network latency while ensuring energy efficiency. By incorporating DRL, this approach uses intelligent agents that learn to make better decisions directly from the experience of interacting with the environment. A testbed setup is used for evaluation and performance analysis. The evaluation of the proposed approach provides better results compared to conventional solutions. Although this solution is performant in terms of ensuring lower latency communication and increasing energy efficiency, some data-related challenges (e.g., privacy) need to be addressed, and FL is certainly the enabling technology to choose.

In \cite{b4}, the authors highlight the problem caused by the uncontrolled development of insecure IoT-based devices and describe an effective NFV infrastructure in combination with emerging technologies that could provide intelligent management and enhanced protection against botnet attacks. In the case of the IoT network, the most common method of intrusion attack is the botnet. It is a large network of interconnected compromised devices running as bots. Typically, a bot attack is accompanied by a Distributed Denial of Service (DDoS) attack, a Man in the Middle (MitM) attack, etc. The authors propose a potential NFV architecture that integrates several emerging technologies: virtual honeynet, cloud computing, and ML. Although the authors describe their proposed solution in detail, it is not implemented and its performance is not evaluated (i.e., tested against real-world known and zero-day botnet attacks).

\begin{table*}[ht]
\caption{Overview of the recent advances on the combination of Network Softwarization, ML, and IoT}
\centering

\begin{tblr}{
  width = \linewidth,
  colspec = {Q[77]Q[65]Q[63]Q[121]Q[71]Q[135]Q[400]},
  row{1} = {c},
  row{2} = {c},
  cell{1}{1} = {r=2}{},
  cell{1}{2} = {c=2}{0.128\linewidth},
  cell{1}{4} = {r=2}{},
  cell{1}{5} = {r=2}{},
  cell{1}{6} = {r=2}{},
  cell{1}{7} = {r=2}{},
  hline{1,3-25} = {-}{},
  hline{2} = {2-3}{},
}
\textbf{Reference} & \textbf{Network Softwarization} &  & \textbf{ML Technique(s)} & \textbf{IoT Networks} & \textbf{Implementation and Evaluation} & \textbf{Objective(s)}\\
 & \textbf{SDN} & \textbf{NFV} &  &  &  & \\
\cite{b17} & Yes & No & No & Yes & Done & Evolving SDN to be adapted to constrained IoT networks.\\
\cite{b11} & Yes & Yes & No & Yes & Done & Providing an energy aware SDN and NFV based architecture for IoT.\\
\cite{b6} & Yes & No & No & Yes & Done & Proposing a mechanism for load balancing routing and virtualization for SDN-enabled IoT.\\
\cite{b18} & Yes & Yes & RL & No & Done & Studying RL-based SFC path selection in SDN and NFV based networks.\\
\cite{b21} & No & Yes & DRL & No & Done & Studying a DRL based framework for online service provisioning in NFV-enabled networks.\\
\cite{b8} & Yes & No & DRL & Yes & Done & Presenting a SDN-based dynamic task scheduling and resource management DRL approach for IoT.\\
\cite{b4} & No & Yes & Yes & Yes & Not done & Proposing an NFV-based scheme for effective protection against bot attacks in AI-enabled IoT.\\
\cite{b3} & Yes & Yes & GNN, DRL & Yes & Done & Proposing a GNN-assisted DRL method for VNF-FG placing in softwarized IoT networks.\\
\cite{b44} & Yes & Yes & Yes & Yes & Done & Proposing an intelligent solution for efficient edge FL communications in SDN and NFV based IoT.
\end{tblr}

\end{table*}

While most existing works assume that services are represented as Service Function Chains (SFCs), which are chains (for reference, we previously reviewed the matter in \cite{b18}), the authors in \cite{b3} consider that network services in IoT networks are more complex and diverse, and therefore a more appropriate representation is VNF Forwarding Graphs (VNF-FGs), which are Directed Acyclic Graphs (DAGs). They note that previous works have failed to exploit this particular graph structure, making them suboptimal or inapplicable to IoT networks. Hence, they investigate the VNF-FG placement problem in dynamic IoT networks where DAG-represented services arrive and depart. In order to fully exploit the graph structures of services and handle the complexity of dynamic IoT networks, the authors combine a Graph Neural Network (GNN) with Deep Reinforcement Learning (DRL) and propose an efficient algorithm for VNF-FG placement, called \textit{Kolin}. Extensive simulation results suggest that \textit{Kolin} outperforms state-of-the-art solutions (e.g., First Fit Dijkstra, Greedy) in terms of system cost, acceptance rate, and computational complexity. 

To address edge FL challenges in large-scale heterogeneous IoT networks (e.g., massive multi-dimensional model update iterations), the authors in \cite{b44} introduce a model of converging SDN and NFV to provide NFV-enabled edge FL aggregation servers to enhance automation and controllability. The proposed solution is ML-based as they use Multi-Agent Deep Q-Networks (MADQNs) to enforce self-learning softwarization. Simulation has been conducted and shows the outperformance of the proposed solution over the reference ones in terms of QoS performance metrics (e.g., PDR, delay, throughput).

Finally, we note that even though some efforts have been made recently to merge ML, network softwarization (mainly SDN and NFV technologies), and IoT, as described above (the number of research papers mainly increased from 2022), more efforts are still needed to achieve effective solutions.

\section{Discussion}
It is evident that academia is fully aware of the potential of SDN, NFV, ML and IoT as key enabling technologies. Each technology has received a lot of attention on its own, and as a result, recent research papers have proposed to merge them. Since network softwarization is essential to ensure efficient IoT networks, and since ML algorithms are more appropriate for softwarized networks, we are moving toward intelligent network softwarization for IoT very soon.

However, after examining the most relevant and recent research works on the topic, which we summarize in Table I, we conclude that the works that exploit the full potential of the cited technologies in conjunction are still sparse \cite{b3} \cite{b44}. We observe that many works that consider network softwarization only consider SDN and omit NFV, even if the combination of both ensures the efficiency of the softwarization. There is an important progress in intelligent network softwarization, but it is not always applicable in IoT constrained networks.

It is also important to point out that few works propose a consistent implementation and evaluation, due to lack of access to the necessary resources (e.g., network datasets, source codes). We concur with all authors who strive to have normalized network datasets to be able to implement solutions and perform coherent evaluations of performance, and thereby encourage the research community to make such resources openly available.

Obviously, the softwarization should be adapted to the peculiarities of IoT networks. Therefore, authors should consider the IoT constraints (e.g., energy limitations) when setting the evaluation metrics. It is important to make consistent comparisons, as we found that authors usually compare their proposals to conventional solutions, not to the smart counterparts.

As future research directions, we suggest focusing on new emerging technologies such as Edge Computing (EC). It allows providing shorter service response times and reducing the cost of processing IoT data in the cloud, while ensuring the offloading of intensive computational tasks from less capable devices to powerful edge servers. It is also interesting to consider a distributed multi-controller softwarization architecture, where multiple controllers coexist in a clustered multi-tier scheme, thus ensuring scalability and reliability while avoiding Single Point of Failure (SPoF). Having such a distributed architecture implies the possibility to propose distributed learning approaches (i.e., FL techniques).

Finally, in order to ensure real implementable solutions in daily life, security should be given the necessary attention as it is a cross-cutting aspect in the intelligent network softwarization of IoT. It is a very large domain and a lot of research could be done towards a self-secured IoT. Intelligence is certainly a driving force in this regard, as evidenced by recent research in this area.

\section{Conclusion}
It is undeniable that there is a critical need to provide agile and scalable IoT networks. Considering the current emerging technologies, intelligent network softwarization is certainly the enabler to address the IoT constraints and challenges. In this paper, we have analyzed the recent advances in this area and conclude that more research efforts should be made towards implementable solutions. 

In future work, we plan to consider these technologies to propose an intelligent network softwarization architecture that addresses multiple IoT challenges. For instance, the proposed general architecture can be oriented to address several IoT challenges such as: energy awareness, improved clustering algorithms, traffic flow control (with respect to QoS requirements), etc.

\section*{Acknowledgments}
The authors would like to express their gratitude to the anonymous reviewers for their constructive and insightful comments.

\end{document}